\documentclass[prb,twocolumn,preprintnumbers,amsmath,amssymb]{revtex4}
\usepackage{graphicx}

\begin{document}

\title{  Phase sensitive two mode squeezing and photon correlations  from  exciton superfluid   }
\author{T. Shi$^{1}$,  Longhua Jiang $^{2}$  and Jinwu Ye$^{2}$ }
\affiliation{$^{1}$Institute of Theoretical Physics, Chinese Academy
of Sciences,
Beijing, 100080, China \\
$^{2}$Department of Physics, The Pennsylvania State University, University
Park, PA, 16802, USA }
\date{\today }

\begin{abstract}
 There have been
experimental and theoretical studies on Photoluminescence (PL) from possible exciton superfluid in
semiconductor electron-hole bilayer systems. However, the PL contains no phase information and no photon correlations, so
it can only lead to suggestive evidences. It is important to identify smoking gun experiments which can lead to convincing evidences.
Here we study  two mode phase sensitive squeezing spectrum and also two photon correlation functions.
We find the emitted photons along all tilted directions are always in a two mode squeezed state between
$ \vec{k} $ and $ - \vec{k} $. There are always two photon bunching, the photon statistics is super-Poissonian.
Observing these unique features by  possible future phase sensitive homodyne experiment and HanburyBrown-Twiss type of experiment
could lead to conclusive evidences of exciton superfluid in these systems.

\end{abstract}

\maketitle

\section{Introduction}

 There have been extensive activities to study the superfluid of two species
 quantum degenerate fermionic gases across the BCS to BEC crossover tuned by the Feshbach resonances \cite{momfer,vortexfer,feshbach}.
 The detection of a sharp peak in the momentum  distribution of fermionic atom pairs gives a suggestive evidence of the superfluid \cite{momfer}.
 However, the most convincing evidence comes from the phase sensitive observation of the vortex lattice across the whole BEC in BCS
 crossover\cite{vortexfer}. In parallel to these achievements in the cold atoms, there have also been extensive experimental search of exciton superfluid \cite{loz} in semiconductor $GaAs/AlGaAs $ electron-hole bilayer system (EHBL) \cite{butov,field}.
 Similar to the quantum degenerate fermionic gases, the EHBL also displays the BEC to BCS crossover tuned by
 the the density of excitons at a fixed interlayer distance\cite{ye}.
 Several features of Photoluminescence (PL) \cite{butov} suggest a
 possible formation of exciton superfluid at low temperature.
 There are also several theoretical work on the PL from the possible exciton superfluid phases \cite{angular,spain,power}.
  Because the PL is a photon density measurement, it  has the following serious limitations:
  (1) It can not detect the quantum nature of emitted photons. (2) It contains no phase information.
  (3) It contains no photon correlations.
  As first pointed out by Glauber \cite{glauber} and others \cite{book1}, it is only in higher-order interference experiments
  involving the interference of photon quadratures or intensities which can distinguish the predictions between classical and quantum theory.
  So the evidence from the PL on possible exciton superfluid is only suggestive.
  Just like in the quantum degenerate fermionic gases, it is very important to perform a phase sensitive
  measurement that can provide a conclusive evidence for the possible exciton superfluid in EHBL. Unfortunately, it is technically impossible
  to rotate the EHBL to look for vortices or vortex lattices.
  In this paper, we show that the two mode phase sensitive measurement which is the interference of photon quadratures in Eqn.\ref{S}
  can provide such a conclusive evidence. We will also study the correlations of the photon intensities
  which is the interference of photon intensities in Eqn.\ref{g2pm}.
  We find that the two mode squeezing spectra  and the two photon correlation
  functions  between $ \vec{k} $ and $ - \vec{k} $ show unique, interesting and rich structures.
  The emitted photons along all tilted directions due to the quasi-particles above the condensate are in a
  two modes squeezed state between in-plane momentum $ \vec{k} $ and $
  - \vec{k} $.  From the two photon correlation functions, we find there are photon
  bunching, the photo-count statistics is super-Poissonian.
  These remarkable features can be used for high precision
  measurements and quantum information processing.
  We also discuses the possible future phase
  sensitive homodyne measurement to detect the two mode squeezing
  spectrum and the HanburyBrown-Twiss type of experiments to detect two
  photon correlations. Observing these unique features by these experiments could lead to conclusive evidences of
  exciton superfluid in these systems.

  The rest of the paper is organized as follows. In Section II, we present the photon-exciton interaction Hamiltonian and the input-output relation between
  incoming and outgoing photons. Then we apply the input-output formalism to study the two mode squeezing between the photons at $ \vec{k} $ and $ - \vec{k} $.
  in section III and the two photon correlations and photon statistics in Section IV.
  We reach conclusions in Sect.V and also present some future open problems.


\section{The Photon-exciton interaction and Input-out formalism }

The total Hamiltonian is the sum of excitonic superfluid part,
photon part and the coupling between the two parts $H_{t}=H_{sf}+H_{ph}+H_{int}$ where :
\begin{eqnarray}
H_{sf} &=&\sum_{\vec{k}}(E_{\vec{k}}^{ex}-\mu )b_{\vec{k}}^{\dagger }b_{\vec{%
k}}+ \frac{1}{2 A} \sum_{\vec{k}\vec{p}\vec{q}}V_{d}(q)b_{\vec{k}-\vec{q}}^{\dagger }b_{%
\vec{p}+\vec{q}}^{\dagger }b_{\vec{p}}b_{\vec{k}}  \nonumber \\
H_{ph} &=&\sum_{k}\omega _{k} a_{k}^{\dagger }a_{k},~~~~
H_{int} =\sum_{k}[ig(k)a_{k}b_{\vec{k}}^{\dagger }+h.c.]
\label{first}
\end{eqnarray}%
 where  $ A $ is the area of the EHBL, the exciton energy $ E_{\vec{k}}^{ex} = \vec{k}^{2}/2M +
 E_{g}-E_{b}$, the photon frequency
 $ \omega_{k}=v_{g}\sqrt{k_{z}^{2}+\vec{k}^{2}}$ where
 $v_{g}=c/\sqrt{\epsilon } $ with $ c $  the light speed in the
 vacuum and $\epsilon \sim 12$  the dielectric constant of $GaAs$,
 $ k=( \vec{k},k_z) $ is the 3 dimensional momentum,
 $V_{d}(\vec{q})$ is the dipole-dipole interaction between the
 excitons \cite{ye}, $V_{d}(\left\vert \vec{r}\right\vert \gg
d)=e^{2}d^{2}/\left\vert \vec{r}\right\vert ^{3}$ and $  V_{d}(q=0)
= \frac{ 2 \pi e^{2} d}{ \epsilon } $  where $ d $ is the interlayer distance leads to a capacitive term for
the density fluctuation \cite{psdw}. The $ g(k) \sim
\vec{\epsilon}_{k\lambda }\cdot \vec{D}_{k} \times L^{-1/2}_{z} $ is
the coupling between the exciton and the photons  where  $ \vec{\epsilon}_{k\lambda } $ is the photon polarization,
$ \vec{D}_{k} $ is the transition dipole moment and $ L_{z}
\rightarrow \infty $ is the normalization length along the $ z $
direction \cite{power}. As emphasized in \cite{power}, the effect of off-resonant pumping in the experiments in \cite{butov}
is just keep the chemical potential $ \mu $ in Eqn.\ref{first} a constant in a stationary state.

\begin{figure}
\includegraphics[width=1.4in]{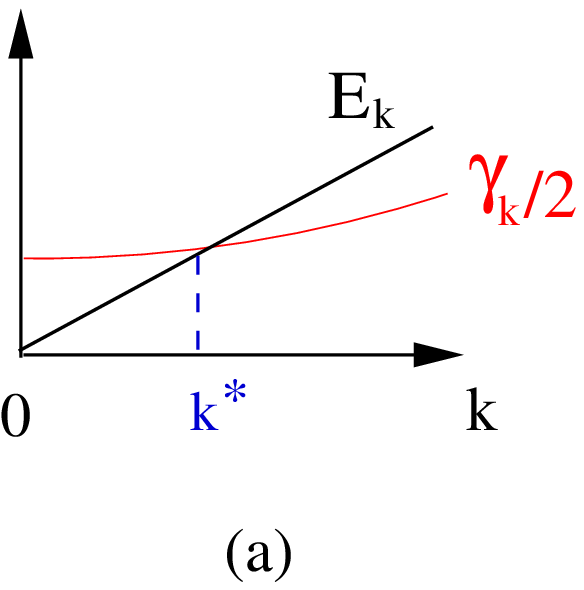}
\hspace{0.3cm}
\includegraphics[width=1.5in,height=1.5in]{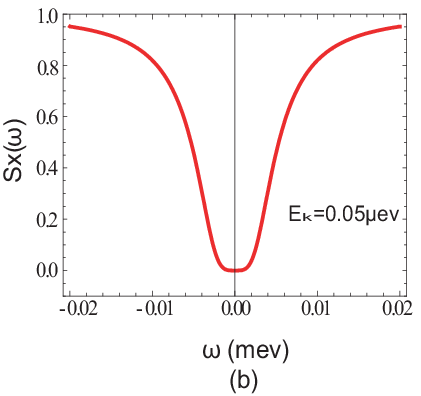}
\caption{(a) The energy spectrum and the decay rate of the exciton
versus in-plane momentum $\vec{k}$ of an exciton superfluid. (b)
The squeezing spectrum  at  a given in-plane momentum $
\vec{k} $ when  $ E(\vec{k}) < \gamma_{\vec{k}}/2 $. The $ n V_{d}(
\vec{k}) $ and $ \gamma_{\vec{k}}/2 $ are fixed at $ 50 \mu eV $ and
$ 0.1 \mu eV $ respectively as used in \cite{power}. They are also
used in all the following figures. There exists only one minimal
when the photon frequency equals to the chemical potential with the
width $ \delta_{1}( \vec{k} ) $ given in the text. Near
the resonance, the squeezing ratio is so close to zero that it can
not be distinguished in the figure. }
\end{figure}

We can apply standard Bogoloubov
approximation to this system. We decompose the exciton operator into
the condensation part and the quantum fluctuation part above the
condensation $ b_{\vec{k}}=\sqrt{N}\delta
_{\vec{k}0}+\tilde{b}_{\vec{k}}$. The excitation spectrum is given
by $ E(\vec{k})=\sqrt{\epsilon _{\vec{k}}[\epsilon
_{\vec{k}}+2\bar{n}V_{d}(\vec{k})]}  $ whose
 $ \vec{k} \rightarrow 0 $ behavior is shown in Fig.1a. We also decompose the interaction Hamiltonian $
H_{int}$ in Eqn.\ref{first} into the coupling to the condensate part
$ H_{int}^{c}=\sum_{k_{z}}[ig(k_{z})(\sqrt{N} + \tilde{b}_{0})
a_{k_{z}}+h.c.] $ and to the quasi-particle part $
H_{int}^{q}=\sum_{k}[ig(k)a_{k}\tilde{b}_{\vec{k}}^{\dagger }+h.c.]
$. The $ \vec{k} = 0 $ part was analyzed in \cite{power}. In this
paper, we focus on the two mode squeezing spectrum and the two
photon correlations between $ \vec{k} $ and $ - \vec{k} $.
The output field $ a_{\vec{k}}^{out}(\omega )$ is
  related to the input field by \cite{online}:
\begin{eqnarray}
a_{\vec{k}}^{out}(\omega ) &=&[-1+\gamma _{\vec{k}}G_{n}(\vec{k},\omega +i%
\frac{\gamma _{k}}{2})]a_{\vec{k}}^{in}(\omega )  \nonumber \\
&&+\gamma _{\vec{k}}G_{a}(\vec{k},\omega +i\frac{\gamma _{k}}{2})a_{-\vec{k}%
}^{in\dagger }(-\omega ),  \label{bout}
\end{eqnarray}%
where the normal Green function $ G_{n}(\vec{k},\omega )
=i\frac{\omega +\epsilon _{\vec{k}}+\bar{n}V_{d}( \vec{k})}{\omega
^{2}-E^{2}(\vec{k})} $ and the anomalous Green function
$G_{a}(\vec{k},\omega ) = \frac{i\bar{n}V_{d}(\vec{k})}{\omega
^{2}-E^{2}( \vec{k})} $ with $ \omega =\omega _{k}-\mu $
\cite{power}. The exciton decay rate in the two Green functions are
$\gamma _{\vec{k}}=D_{\vec{k}}(\mu )\left\vert g_{\vec{k}}(\omega
_{k}=\mu )\right\vert ^{2} $ which is independent of $ L_{z} $
\cite{power}, so is an experimentally measurable quantity.  Just
from the rotational invariance, we can conclude that $ \gamma
_{\vec{k}} \sim const.+ |\vec{k}|^{2} $ as $ \vec{k} \rightarrow 0 $
as shown in Fig.1a.

\section{The two modes squeezing between $ \vec{k} $ and $ -\vec{k} $. }

 Eqn.\ref{bout} suggests that it is convenient to define
$A_{\vec{k},\pm }^{out}(\omega ) =[a_{\vec{k}}^{out}(\omega )\pm
a_{-\vec{k}}^{out}(\omega )]/ \sqrt{2}$ and $A_{\vec{k},\pm
}^{in}(\omega )=[a_{\vec{k}}^{in}(\omega )\pm
a_{-\vec{k}}^{in}(\omega )]/\sqrt{2}$. Then the position and
momentum ( quadrature phase ) operators of the output field
can be defined by:
\begin{eqnarray}
X_{\pm } &=&A_{\vec{k},\pm }^{out}(\omega )e^{i\phi _{\pm }(\omega )}+A_{%
\vec{k},\pm }^{out\dagger }(-\omega )e^{-i\phi _{\pm }(-\omega )}  \nonumber
\\
iY_{\pm } &=& A_{\vec{k},\pm }^{out}(\omega )e^{i\phi _{\pm }(\omega )}-A_{%
\vec{k},\pm }^{out\dagger }(-\omega )e^{-i\phi _{\pm }(-\omega )}
\label{rotate}
\end{eqnarray}%
The squeezing spectra \cite{book1} which measure the fluctuation of the
canonical position and momentum are defined by%
\begin{eqnarray}
S_{X_{\pm }}(\omega ) &= &\left\langle X_{\pm }(\omega )X_{\pm }(-\omega
)\right\rangle _{in}   \nonumber \\
S_{Y_{\pm }}(\omega ) & = &\left\langle Y_{\pm }(\omega )Y_{\pm }(-\omega
)\right\rangle _{in}.  \label{S}
\end{eqnarray}%
where the in-state is the initial zero photon state $ |in \rangle=
|BEC \rangle |0 \rangle $. For notational conveniences, we set $\phi
_{-}(\omega )=\pi /2+\phi _{+}(\omega )$ and just set $\phi
_{+}(\omega )\equiv \phi (\omega )$. Then
we find $S_{X_{+}}(\omega )=S_{X_{-}}(\omega )=S_{X}(\omega )$ and $%
S_{Y_{+}}(\omega )=S_{Y_{-}}(\omega )=S_{Y}(\omega )$. The phase $\phi
(\omega )$ is chosen to achieve the largest possible squeezing, namely, by
setting $\partial S_{X}(\omega )/\partial \omega =0$ which leads to:
\begin{equation}
\cos 2\phi (\omega )=\frac{\gamma _{\vec{k}}(\epsilon _{\vec{k}}+\bar{n}%
V_{d}(\vec{k}))}{\sqrt{\Omega ^{2}(\omega )+\gamma _{\vec{k}}^{2}E^{2}(\vec{k%
})+(\bar{n}V_{d}(\vec{k})\gamma _{\vec{k}})^{2}}},  \label{angle}
\end{equation}%
where $\Omega (\omega )=\omega ^{2}-E^{2}(\vec{k})+\gamma
_{\vec{k}}^{2}/4$.

Substituting Eqns.\ref{bout} and \ref{rotate} into Eq.\ref{S}
leads to%
\begin{eqnarray}
S_{X}(\omega ) &=&1-\frac{2\gamma _{\vec{k}}\bar{n}V_{d}(\vec{k})}{\mathcal{N%
}(\omega )+\gamma _{\vec{k}}\bar{n}V_{d}(\vec{k})}  \nonumber \\
S_{Y}(\omega ) &=&1+\frac{2\gamma _{\vec{k}}\bar{n}V_{d}(\vec{k})}{\mathcal{N%
}(\omega )-\gamma _{\vec{k}}\bar{n}V_{d}(\vec{k})}  \label{squ}
\end{eqnarray}%
where $\mathcal{N}(\omega )=\sqrt{\Omega ^{2}(\omega )+\gamma _{\vec{k}%
}^{2}E^{2}(\vec{k})+(\bar{n}V_{d}(\vec{k})\gamma _{\vec{k}})^{2}}$. This equation leads to:
\begin{equation}
S_{X}(\omega )S_{Y}(\omega )=1.
\label{squeeze}
\end{equation}%
which shows that for a given in-plane momentum $\vec{k}$ and a given
photon frequency $\omega=\omega _{k}-\mu $, there always exists a two mode squeezing state which can be
decomposed into two squeezed states along two normal angles: one
squeezed along the angle $\phi (\omega )$ and the other along the angle $%
\phi (\omega )+\pi /2$ in the quadrature phase space ($X,Y$).
 Now we discuss the over-damping $ k< k^{\ast}, E(\vec{k}) < \gamma _{\vec{k}}/2 $ case
 and the under-damping $ k > k^{\ast}, E(\vec{k})> \gamma _{\vec{k}}/2 $ case respectively.

{\sl (1) Low momentum regime $|\vec{k}| < k^{\ast }$:
$E(\vec{k})<\gamma _{\vec{k}}/2$. }

 From Eqn.\ref{squ}, we can see that the maximum squeezing happens at
$\omega _{\min }=0$ which means at $\omega _{k}=\mu $:
\begin{eqnarray}
S_{X}(\vec{k},\omega =0) &=&1-\frac{2\gamma _{\vec{k}}\bar{n}V_{d}(\vec{k})}{%
\mathcal{N}(0)+\bar{n}V_{d}(\vec{k})\gamma _{\vec{k}}}  \nonumber \\
\cos 2\phi (\vec{k},\omega =0) &=&\frac{\gamma _{\vec{k}}(\epsilon _{\vec{k}%
}+\bar{n}V_{d}(\vec{k}))}{\mathcal{N}(0)}  \label{S2}
\end{eqnarray}%
where $\mathcal{N}(0)\equiv \mathcal{N}(\omega =0)=\sqrt{[E^{2}(\vec{k}%
)+\gamma _{\vec{k}}^{2}/4]^{2}+(\bar{n}V_{d}(\vec{k})\gamma _{\vec{k}})^{2}}$%
which is defined below Eqn.\ref{squ}.  In sharp contrast
to the large momentum regime $ E(\vec{k})
> \gamma _{\vec{k}}/2$ to be discussed in the following, the resonance position $\omega
_{k}=\mu $ is independent of the value of $ \vec{k} $, this is
because the quasiparticle is not even well defined in the low momentum regime\cite{power}.
 The $\omega $ dependence of $S_{X}(\omega )$ in Eqn.\ref{squ} is drawn in Fig.1b. The line width
of the single peak in Fig.1b is $
\delta_{1}(\vec{k})=2\sqrt{E^{2}(\vec{k})-\frac{\gamma _{\vec{k}}^{2}}{4}+O_{%
\vec{k}}} $
where $ O_{\vec{k}}=\sqrt{4\mathcal{N}(0)[\mathcal{N}(0)+\bar{n}V_{d}(\vec{k})\gamma
_{\vec{k}}]-\gamma _{\vec{k}}^{2}E^{2}(\vec{k})} $.

\begin{figure}
\includegraphics[width=3.0in,height=1.5in]{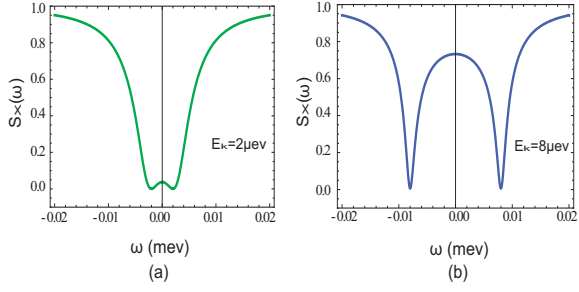}
\caption{The squeezing spectrum  at  a given in-plane momentum $
\vec{k} $ when  $ E(\vec{k}) > \gamma _{\vec{k}}/2 $. There exist
two minima in the spectrum when the photon frequency resonate with
the well defined quasi-particles. Near the resonance, the squeezing
ratio is so close to zero that it can not be distinguished in the
figure. (a) When $ E(\vec{k})=2 \mu eV $, the two peaks are still not
clearly separated.  (b) When $E(\vec{k})=8 \mu eV \gg \gamma
_{\vec{k}}/2$, the quasi-particles are well defined which lead to
the two well defined resonances with the width $ \delta_{2}( \vec{k}
) $ given in the text.}
\end{figure}

{\sl (2) Large momentum regime $ k > k^{\ast} $:  $ E(\vec{k})> \gamma_{\vec{k}}/2$.}

 From Eqn.\ref{squ}, we can see that the maximum squeezing happens at the two resonance frequencies $\omega
_{k}=\mu \pm \lbrack E^{2}(\vec{k})-\gamma _{\vec{k}}^{2}/4]^{1/2}$ where
\begin{eqnarray}
S_{X}(\vec{k},\omega _{\min }) & = & ( \frac{ \epsilon_{\vec{k}}}{ E(\vec{k}%
) })^{2} = \frac{ \hbar
k^{2} }{ \hbar^{2} k^{2} + 4 M \bar{n}V_{d}(\vec{k}) }  \nonumber \\
\cos 2 \phi(\vec{k},\omega _{\min }) & = & 1  \label{S1}
\end{eqnarray}
In this case, $\phi (\vec{k}, \omega _{\min })=0$. In sharp contrast
to the low momentum regime discussed above, the resonance
positions depend on $ E (\vec{k}) $, this is because the
quasiparticle is well defined in the large momentum regime only
\cite{power}. From Eqn.\ref{S1}, we can see that increasing the
exciton mass, the density, especially {\sl the exciton dipole-dipole
interaction } will all benefit the squeezing at the two resonances.


The $\omega $ dependence of $S_{X}(\omega )$ in Eqn.\ref{squ} is
drawn in Fig.2. When $E(\vec{k})>(Q_{\vec{k}}+\sqrt{1+Q_{\vec{k}}})\gamma _{\vec{k}%
}/2 $, the line width of the each peak in Fig.2 is
$ \delta_{2}(\vec{k}) =\sqrt{E^{2}(\vec{k})-\frac{\gamma
_{\vec{k}}^{2}}{4}+\gamma _{\vec{k}}Q_{\vec{k}}E(\vec{k})} -\sqrt{E^{2}(\vec{k})-\frac{\gamma
_{\vec{k}}^{2}}{4}-\gamma _{\vec{k}}Q_{ \vec{k}}E(\vec{k})}
$ where $ Q_{\vec{k}}=\sqrt{3+\frac{4\bar{n}V_{d}(\vec{k})}{\epsilon
_{\vec{k}}}} $. It is easy to see that $ \delta_{2} \sim
\gamma_{\vec{k}} Q_{\vec{k}} $ which is equal to the exciton decay
rate $ \gamma_{\vec{k}} $ multiplied by a prefactor $ Q_{\vec{k}} $.
When $E(\vec{k})<(Q_{\vec{k}}+\sqrt{1+Q_{\vec{k}}})\gamma
_{\vec{k}}/2$, the two peaks are too close to be distinguished.
It is important to observe that the two widths $ \delta_{1}( \vec{k}
) $  and $ \delta_{2}( \vec{k}) $ not only depend on $ \gamma _{\vec{k}} $,
but also the interaction $ \bar{n}V_{d}(\vec{k}) $. This is in sharp
contrast to the widths in the ARPS and EDC in\cite{power} which only depend on $ \gamma _{\vec{k}} $.

The angle dependence of both $E(\vec{k})>\gamma _{\vec{k}}/2$ and $
E(\vec{k})< \gamma _{\vec{k}}/2$ are drawn in the same plot Fig.3a
for comparison. Both the squeezing spectrum in Eqn.\ref{squ}  and
the rotated phase $ \phi $ in Eqn.\ref{angle} can be measured by
phase sensitive homodyne detections \cite{book1}.

\begin{figure}
\includegraphics[width=3.2in]{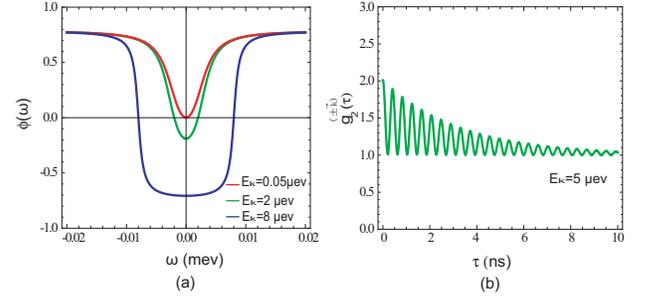}
\caption{ (a) The squeezing angle dependence on the frequency when
$E(\vec{k})< \gamma _{\vec{k}}/2$  corresponding to Fig.1b and $E(\vec{k})> \gamma
_{\vec{k}}/2$ corresponding to Fig.2a,2b.  When $E(\vec{k})< \gamma _{\vec{k}}/2$ the squeezing
angle is always non-zero. Near the resonance, the angle is so close
to zero that it can not be distinguished in the figure. Only when
$E(\vec{k})> \gamma _{\vec{k}}/2$ and the photon frequency resonate
with the well defined quasi-particles, the squeezing angle is zero.
Away from the resonance, the angle becomes negative. (b)The two
photon correlation functions between $ \vec{k} $ and $ -\vec{k} $
against the delay time $ \tau $.}
\end{figure}

\section{ The two photons correlation functions and photon statistics. }

 The quantum statistic properties of emitted photons can be extracted
from two photon correlation functions\cite{glauber}. The normalized
second order correlation functions of the output field for the two
modes at $ \vec{k} $ and $ -\vec{k} $ are
\begin{equation}
g_{2}^{(\vec{k})}(\tau )=\frac{\left\langle a_{\vec{k}}^{out\dagger }(t)a_{%
\vec{k}}^{out\dagger }(t+\tau )a_{\vec{k}}^{out}(t+\tau )a_{\vec{k}%
}^{out}(t)\right\rangle_{in} }{\left\vert G_{1}(0)\right\vert ^{2}}
\label{g2}
\end{equation}%
and%
\begin{equation}
g_{2}^{(\pm \vec{k} )}(\tau )=\frac{\left\langle
a_{\vec{k}}^{out\dagger }(t+\tau
)a_{\vec{k}}^{out}(t+\tau )a_{-\vec{k}}^{out\dagger }(t)a_{-\vec{k}%
}^{out}(t)\right\rangle_{in} }{\left\vert G_{1}(0)\right\vert ^{2}}
\label{g2pm}
\end{equation}%
 where the $ G_{1}( \tau ) = \langle
  a_{\vec{k}}^{out\dagger }(t+\tau ) a_{\vec{k}}^{out}(t)\rangle_{in}$
is the single photon correlation function \cite{power}. The second
order correlation function $g_{2}^{(\pm \vec{k} )}(\tau )$
determines the probability of detecting $n_{-\vec{k} }$ photons with
momentum $-\vec{k} $ at time $t$ and detecting $n_{\vec{k}}$ photons
with momentum $ \vec{k} $ at time $t+\tau $. By using
Eqn.\ref{bout}, we find
\begin{eqnarray}
g_{2}^{(\vec{k})}(\tau )& = & 1+e^{-\gamma _{\vec{k}}\tau }[\cos
(E(\vec{k})\tau )+\frac{\gamma _{\vec{k}}}{2E(\vec{k})}\sin
(E(\vec{k})\tau )]^{2} \nonumber \\
g_{2}^{(\pm \vec{k} )}(\tau ) & = &  g_{2}^{(\vec{k})}(\tau ) +
  e^{-\gamma _{\vec{k}}\tau }\frac{E^{2}(\vec{k})+%
  \frac{\gamma _{\vec{k}}^{2}}{4}}{\bar{n}^{2}V_{d}^{2}(k)}
 \label{gpm}
\end{eqnarray}
It turns out that the second correlation functions are independent
of the relation between $E(\vec{k})$ and $\gamma _{\vec{k}}/2$.
We only draw $ g_{2}^{(\pm \vec{k} )}(\tau ) $ in the Fig.3b.
When $\tau =0$ the two photon correlation function are $g_{2}^{(\vec{k}%
)}(0)=2$, so just the mode $ \vec{k} $ alone behaves like a chaotic
light. This is expected because the entanglement is only between $ -
\vec{k} $ and $ \vec{k} $. In fact,
$ g_{2}^{(\pm k )}( 0 )=2+\frac{E^{2}(\vec{k})+\frac{\gamma _{\vec{k}}^{2}}{4}%
}{\bar{n}^{2}V_{d}^{2}(\vec{k})} > g_{2}^{(\vec{k})}(0)=2 $.
  So it violates the classical Cauchy-Schwarz inequality \cite{book1} which is
  completely due to the {\sl quantum nature } of the two mode squeezing
  between $ \vec{k} $ and $ - \vec{k} $.

  From Fig.3b, we can see that the two photon
  correlation function decrease as time interval $\tau $ increases
  which suggests quantum nature of the emitted photons is photon
  bunching and the photo-count statistics is super-Poissonian.
  It is easy to see that the envelope decaying function is given by
  the exciton decay rate $ \gamma_{\vec{k}} $ shown in Fig.1b, while the oscillation
  within the envelope function is given by the Bogoliubov
  quasi-particle energy $  E( \vec{k} ) $ shown also in Fig.1b. The $ g_{2}^{(\pm \vec{k} )}(\tau ) $ can
  be measured by  HanburyBrown-Twiss type of experiment \cite{book1} where one can extract
  both $  E( \vec{k} ) $ and $ \gamma_{\vec{k}} $.

\section{Conclusions and Perspectives }

  In conventional non-linear quantum optics, the generation of
squeezed lights requires an action of a strong classical pump and a large non-linear susceptibility $%
\chi ^{(2)}$. The first observation of squeezed lights was achieved
in  non-degenerate four-wave mixing in atomic sodium in 1985
\cite{fourwave}. Here in EHBL, the generation of the two mode
squeezed photon is due to a complete different and new mechanism:
the anomalous Green function of Bogoliubov quasiparticle which is
non-zero only in the excitonic superfluid state. The very important two mode squeezing result Eqn.\ref{squeeze} is robust
against any microscopic details such as the interlayer distance $ d $, exciton density $ \bar{n} $, exciton dipole-dipole interaction $ V_{d}(q) $ and
the exciton decay rate $ \gamma_{\vec{k}} $.
The applications of the squeezed state include (1) the very high
precision measurement by using the quadrature with reduced quantum
fluctuations such as the $ X $ quadrature in the Fig.1b and Fig.2 where
the squeeze  factor reaches very close to $ 0 $ at the resonances
(2) the non-local quantum entanglement between the two twin photons
at $ \vec{k} $ and $ - \vec{k} $ can be useful for many quantum
information processes. (3) detection of possible gravitational waves \cite{grav}.
All these various salient features of the phase sensitive two mode squeezing spectra and the two photon correlation functions
along normal and titled directions  studied in this paper can map out
completely  and unambiguously  the nature of quantum
phases of excitons in EHBL such as the ground state and the
quasi-particle excitations above the ground state.
Both the squeezing spectrum in Fig.1b, Fig.2  and
the rotated phase $ \phi $ in Fig.3a can be measured by
phase sensitive homodyne detections.  The two photon correlation functions in Fig.3b can
be measured by  HanburyBrown-Twiss type of experiments. It is important to perform these these new experiments in the future
to search for the most  convincing evidences for the existence of exciton superfluid in the EHBL.
The results achieved in this paper should also shed lights on how photons interact with cold atoms \cite{momfer,vortexfer,feshbach}.

    It was well known that a 2D superfluid at any finite temperature was given by the Kosterlitz-Thouless (KT) physics.
    Namely, at any non-zero temperature, there is no real BEC ( no real symmetry breaking ), only algebraic order where
    the correlation function decays algebraically. But the gapless superfluid mode with a finite superfluid density
    survives at finite temperatures upto the KT transition temperature. So the the condensate at $ \vec{k}=0 $ at $ T=0 $
    will disappear at any finite $ T $, so the properties of the photons emitted along the normal direction at $ T=0 $
    studied in Ref. 10 need to be re-investigated at any finite $ T $.
    This manuscript focused on the interaction between the photons and the Bogoliubov mode at tilted directions
    $ \vec{k} \neq 0 $, because the  Bogoliubov mode is just the gapless superfluid mode which survives at finite temperatures until
    to the KT transition temperature, so we expect the results achieved in this manuscript will also survive at finite temperatures.
    How it will change near the KT transition temperature is an open problem to be discussed in a future publication.
    In all the experiments \cite{butov},  the excitons are confined inside a trap , so there still could be
    a real BEC at finite temperature inside a trap. So the effects of trap also will also be investigated in a future publication.



 {\bf ACKNOWLEDGEMENTS }

We are very grateful for C. P. Sun for very helpful discussions and encouragements throughout the
preparation through this manuscript. J. Ye's research at KITP was supported in part by the NSF under
grant No. PHY-0551164, at KITP-C was supported by
the Project of Knowledge Innovation Program (PKIP) of Chinese
Academy of Sciences. J.Ye thank Fuchun Zhang, Jason Ho, A. V. Balatsky, Jiangqian You, Yan Chen
and Han Pu for their hospitalities during
his visit at Hong Kong university, Ohio State
university, LANL, Fudan university and Rice university.

\end{document}